\documentclass[letter,twocolumn]{jpsj3}
\usepackage{txfonts}
\usepackage{bm}
\usepackage{color}
\title{Slow Electronic Dynamics in the Paramagnetic State of UTe$_2$}

\author{Yo Tokunaga$^1$\thanks{E-mail: tokunaga.yo@jaea.go.jp}, Hironori Sakai$^1$, Shinsaku Kambe$^1$, Yoshinori Haga$^1$,Yoshifumi Tokiwa$^1$, Petr Opletal$^1$, Hiroki Fujibayashi$^2$,
Katsuki Kinjo$^2$, Shunsaku Kitagawa$^2$, Kenji Ishida$^2$, Ai Nakamura$^3$, Yusei Shimizu$^3$, Yoshiya Homma$^3$, Dexin Li$^3$, Fuminori Honda$^{3,4}$, and Dai Aoki$^{3,5}$}

\inst{$^1$ASRC, Japan Atomic Energy Agency, Tokai-mura, Ibaraki 319-1195, Japan  \\
$^2$Department of Physics, Kyoto University, Kyoto 606-8502, Japan  \\
$^3$IMR, Tohoku University, Oarai, Ibaraki, 311-1313, Japan  \\
$^4$Central Institute of Radioisotope Science and Safety, Kyushu University, Fukuoka 819-0395, Japan  \\
$^5$University Grenoble Alpes, CEA, IRIG-PHELIQS, F-38000 Grenoble, France} 

\abst{
$^{125}$Te NMR experiments in field ($H$) applied along the easy magnetization axis (the $a$-axis) revealed slow electronic dynamics developing in the paramagnetic state of UTe$_2$. The observed slow fluctuations are concerned with a successive growth of long-range electronic correlations below 30$-$40 K, where the spin susceptibility along the hard magnetization axis (the $b$-axis) shows a broad maximum.
The experiments also imply that tiny amounts of disorder or defects locally disturb the long-range electronic correlations and develop an inhomogeneous electronic state at low temperatures, leading to a low temperature upturn observed in the bulk-susceptibility in $H\|a$.
We suggest that UTe$_2$ would be located on the paramagnetic side near an electronic phase boundary, where either magnetic or Fermi-surface instability would be the origin of the characteristic fluctuations.
}

\begin{document}
\maketitle

A recently discovered heavy-fermion superconductor UTe$_2 $\cite{Ran2019} 
attracted particular attention,  because of the strong possibility of spin-triplet Cooper pairing, which
is a natural candidate for topological superconductivity (SC) in a bulk material \cite{Read,Ivanov,Fu,Sato}.
The formation of the spin-triplet pairing was experimentally suggested from
tiny decrease in NMR Knight shift below the SC transition temperature $T_{\rm SC}\sim 1.6$ K \cite{Nakamine2019,NakaminePRB2021,NakamineJPSJ2021}, and an exceptionally large upper critical field $H_{c2}$ far exceeding the Pauli-limiting field \cite{Ran2019,Aoki2019,Knebel2019,Knebel2020,RanNature}.
The spin-triplet pairing was further supported by the later discovery of a field-induced reentrant SC in high magnetic field ($H$) over 40 T \cite{RanNature,Knebel2019,Knebel2020,KnafoComm} and multiple SC phases under applied pressure \cite{Aoki2020,Daniel2019}.

UTe$_2$ was initially proposed to be located at the paramagnetic end of the uranium-based ferromagnetic (FM) superconductor series, where SC is mediated by FM fluctuations \cite{Ran2019}.
This was based on the scaling analysis of the bulk spin-susceptibility below 9 K \cite{Ran2019}, and the temperature ($T$) dependence of $\mu$SR relaxation \cite{Sunder}.
However, a key difference separating UTe$_2$ from the other FM superconductors  
is the absence of a long-range magnetic order in the ground state, and as a consequence, the nature of spin fluctuations has been found to be much more complex than originally thought.
NMR $1/T_2$ measurements implied the development of slow and longitudinal fluctuations around 20 K in $H\|a$ \cite{Tokunaga2019}.
The zero-field $\mu$SR experiment also observed similar fluctuations, but only below 5 K \cite{Sunder}. These fluctuations were attributed to possible FM instability of the material. 
On the other hand, more recent neutron scattering experiments detected antiferromagnetic (AFM), incommensurate fluctuations with a $q$-vector of (0, 0.57, 0) \cite{Duan2020,Knafo2021}.
The AFM fluctuations are developing from higher $T$, and saturated below about 12 K \cite{Knafo2021}, similar to the $T$-dependence of NMR $1/T_1T$ \cite{Tokunaga2019}.
Because of the puzzling results from the different probes, a unified picture of fluctuations has not yet emerged for UTe$_2$.

\begin{figure}[tb]
\begin{center}
\includegraphics[width=8.5 cm,keepaspectratio]{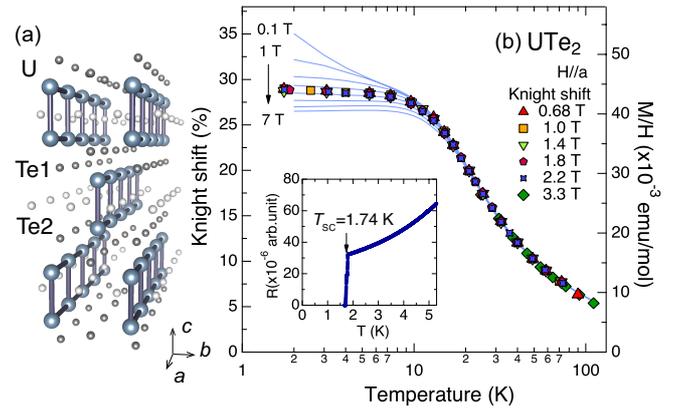} 
\caption{(Color online) (a) The crystal structure of UTe$_2$, where the ladder structure of U atoms is emphasized. (b) Temperature and field dependences of the NMR Knight shift $K_{\rm a}(T,H)$ and the bulk-susceptibility, $\chi_{\rm a}(T,H)$.
The $\chi_{\rm a}$ was measured in fields H$=0.1$ T and 1 - 7 T in 1 T step. 
The inset shows the resistivity data of our single crystal.
}
\end{center}
\label{f1}
\vspace{-7mm}
\end{figure}

In this paper, we report $^{125}$Te-NMR experiments performed on a 
$^{125}$Te enriched (99\%) single crystal of UTe$_2$.
The enrichment of $^{125}$Te largely enhanced NMR signal intensity and S/N ratio, allowing us to extend NMR studies to lower $H$ and $T$ than a previous report \cite{Tokunaga2019}. 
The single crystal was grown using the chemical vapor transport method with iodine as transport agent \cite{Aoki2019}. As seen in the inset of Fig.\,1(b), the resistivity data of our single crystal shows a SC transition at $T_{\rm SC}=$1.74 K defined by the midpoint of the resistivity drop; a little lower than the highest $T_{\rm SC}$ of $\sim$2 K reported recently \cite{Rosa}.
In the orthorhombic structure of UTe$_2$ (space group No. 71, $Immm$, $D_{2h}^{25}$),  U atoms in equivalent sites form a two-leg ladder structure with legs along the $a$-axis and rungs along the $c$-axis, with the inversion center located between the two U atoms, as displayed in Fig.\,1(a) \cite{Hutanu,Knafo2021}.
On the other hand, Te atoms surrounding the ladder have two crystallographically inequivalent sites in a unit cell, so that $^{125}$Te NMR spectrum consists of two distinct peaks arising from the two inequivalent sites in $H \| a$ \cite{Tokunaga2019}. Although we measured both peaks, there was no qualitative difference in the NMR results; here we focus on the data obtained at the lower frequency peak within two peaks.

NMR spectra were measured by recording integrated spin-echo intensities while sweeping frequency at a fixed $H$. 
The NMR
(Knight) shift was derived from the peak position of the NMR spectrum, where Cu NMR signals from metallic copper were used as markers for field calibrations.
The nuclear spin-spin relaxation rate ($1/T_2$) was obtained by measuring the decay of the spin-echo intensity on the peak position. The decay was found to fit an exponential function, $M(2\tau)\propto \exp(-2\tau/T_2)$ as reported previously\cite{Tokunaga2019}, where $\tau$ is the time interval between the first and second pulses in the NMR spin-echo sequence.

\begin{figure}[tb]
\begin{center}
\includegraphics[width=7.3 cm,keepaspectratio]{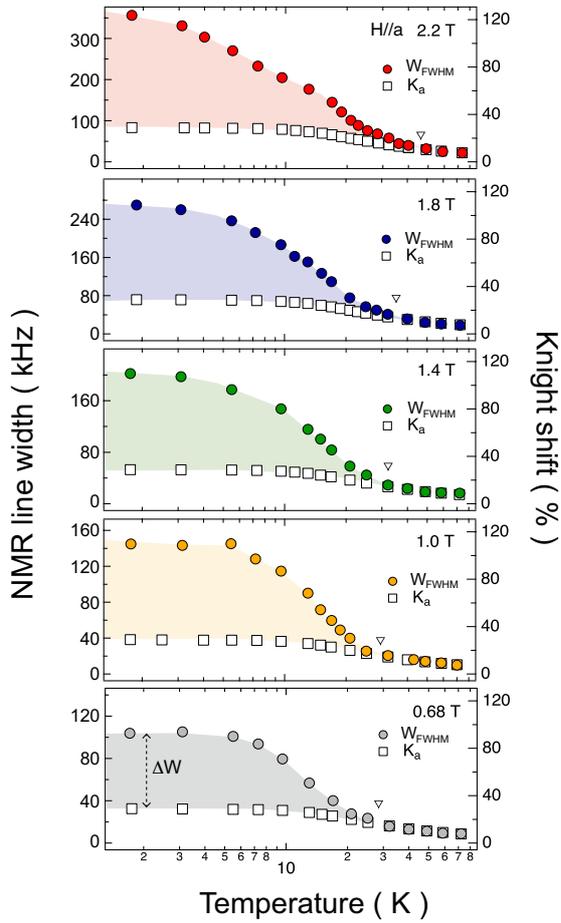}  
\caption{(Color online) Comparison of the $T$ dependences of NMR line width (FWHM) and Knight  shift measured at several different $H$ in $H||a$. The down triangles indicate the temperature below which additional NMR line broadening emerges. }
\end{center}
\vspace{-5mm}
\end{figure}

\begin{figure}[tb]
\begin{center}
\includegraphics[width=7.3 cm,keepaspectratio]{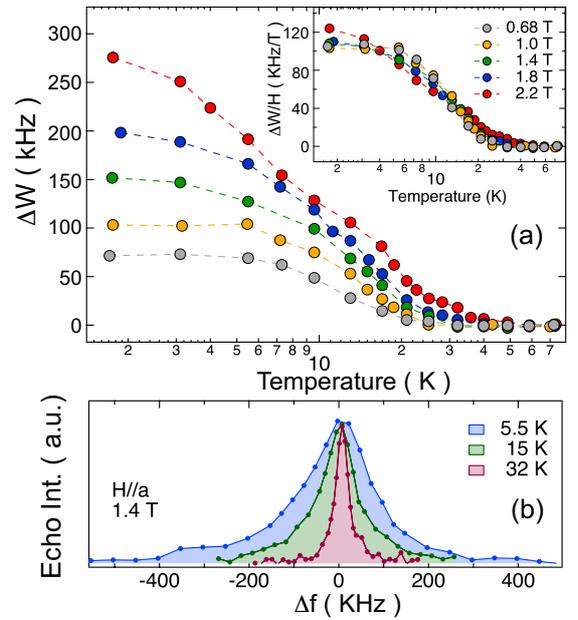} 
\caption{(Color online) 
(a) Temperature dependence of $\Delta W$ in different fields. The inset shows the $T$ dependence of $\Delta W/H$.
(b) Comparison of NMR spectra obtained at $T$=5.5 K, 15 K and 32 K in $H$=1.4 T. 
The horizontal axis $\Delta f=f_{\rm meas.}-f_{\rm peak}$ is the frequency difference from
the peak frequency. The spectrum intensity is normalized by the intensity at the peak position.  }
 \end{center}
\vspace{-5mm}
\end{figure}

Figure 1(b) shows the $T$ and $H$ dependences of bulk spin-susceptibility, $\chi_a$ and the Knight shift $K_{\rm a}$ in $H \| a$. 
For UTe$_2$, the susceptibility, and thus the Knight shift, exhibit Ising anisotropy with the $a$-axis as the magnetic easy axis, and the $\chi_a$ follows a Curie Weiss behavior with a negative Weiss temperature of $\theta=-60$ K above 150 K \cite{Ran2019,KnafoComm,Ikeda}.  
With further decreasing $T$, the $\chi_a$ starts to increase faster than the Curie Weiss behavior, and tends to saturate below 20 K.
The $\chi_a$ becomes strongly $H$ dependent below 20 K, and has an additional upturn below 10 K at lower $H$ ($\lesssim2$ T) \cite{Ran2019,Miyake2019,KnafoComm,Rosa}.
In contrast, we found that the $K_{\rm a}$ is totally $H$ independent over the whole $T$ region. Consequently the linear scaling obtained at higher $T$ between the $K_{\rm a}(T)$ and $\chi_{\rm a}(T)$ does not work below 20 K.
Below 10 K, the $K_{\rm a}$ becomes almost $T$ independent, and thus never reproduce the low-$T$ upturn observed in the $\chi_a(T)$.
 
In general, the discrepancy in the $T$ dependence between the bulk-$\chi$ and the NMR $K$ implies the presence of foreign impurity phases in measured samples;  
the $K$ is insensitive to the small amounts of foreign phases, 
whereas the bulk-$\chi$ involves contributions from both the main and foreign phases.
However, this is probably not the case for UTe$_2$, because the low-$T$ anomaly in the $\chi_{\rm a}(T)$ has been commonly observed in single crystals grown by different groups \cite{Ran2019,Miyake2019,KnafoComm,Rosa}. In addition, the anomaly is strongly field-orientation dependent (not observed in $H||b$ and $||c$), and suppressed rapidly under applied pressure \cite{Li2021}.
Furthermore, we have observed a distinct anomaly in the $T$  dependence of NMR line width and $1/T_2$ in the same $T$ region, as described below. 
These imply that the low-$T$ anomaly in the $\chi_a(T)$ does not arise from foreign impurity phases, but is rather connected to electronic properties of UTe$_2$ \cite{UTe2Review}.

Figure 2 shows the $T$ dependence of the NMR line width (the full width at half maximum; $W_{\rm FWHM}$) measured in several different $H$.
In the figure, we also plotted the $K_{\rm a}(T)$ obtained in the same $H$.
The  $W_{\rm FWHM}$ increases with decreasing $T$. 
At higher-$T$, it entirely scales with the $K_{\rm a}$, as expected when the broadening simply arises from a small distribution of the $K_{\rm a}$. However, below about 30-40 K, the  $W_{\rm FWHM}$ starts to increase much faster than that of the $K_{\rm a}$,
as if either the onset of spontaneous magnetic ordering or the sudden growth of electronic inhomogeneity took place at low $T$. 
The onset temperature of the broadening shifts to higher $T$ in higher $H$, as shown by the open triangles in Fig.\,2.
In Fig.\,3 (a), we extracted  the additional broadening component as $\Delta W= W_{\rm FWHM} - W_{\rm \Delta K}$, where $W_{\rm \Delta K}=\alpha K_{\rm a}$ with a scaling coefficient $\alpha$ determined above 30$-$40 K. 
Although the absolute values of $\Delta W$ are enhanced with increasing $H$, we found that all the data fall roughly on a single curve after dividing the values by $H$, as seen in the inset of Fig.\,3 (a). 
This indicates that the $\Delta W$ is nearly proportional to $H$, i.e., $\Delta W \rightarrow 0$ at $H \rightarrow 0$. The $\Delta W$ is thus field-induced, not spontaneous one. 
 
 Figure 3 (b) displays typical evolution of the NMR line profile, which corresponds to the histogram of local fields associated with the electronic inhomogeneity.
As mentioned above, the spectra show a distinct broadening below 30$-$40 K. However, it holds nearly symmetrical and structureless profile, indicating continuous distribution of local fields in magnitude.
Such a structureless profile will deny any homogeneous, long-range magnetic orderings, neither with commensurate nor incommensurate wave vector in the 1D system. Interestingly, a similar broadening has been observed in low-dimensional correlated magnets with quenched disorder, such as x-ray-irradiated organic materials \cite{Itou,Yamamoto,Urai}, and impurity-doped spin-ladder \cite{Fujiwara,Ohsugi,Alexander,Casola} and high-$T_{\rm SC}$ cuprates \cite{Bobroff,Julien,Ouazi,Alloul}. 
In these materials, defects or impurities locally disturb long-range electronic correlations, induce local moments (magnetic clusters) at their neighbor sites, and develop inhomogeneous spin susceptibility. This results in discrepancy between the bulk-$\chi$ and the NMR $K$ at low $T$.
 
\begin{figure}[tb]
\begin{center}
\includegraphics[width=7.0 cm,keepaspectratio]{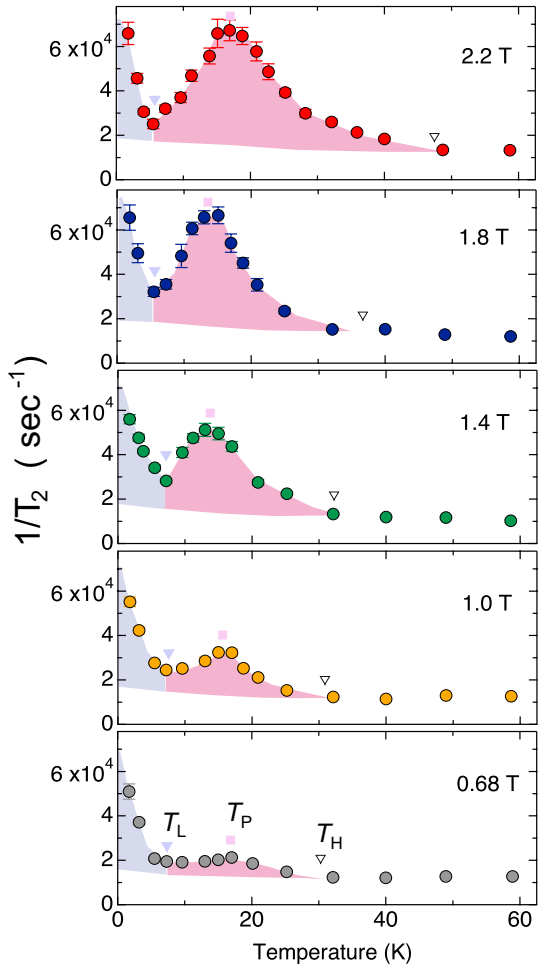} 
\caption{(Color online) The $T$ dependence of $1/T_2$ measured at several different $H$ in $H\|a$. The $1/T_2$ starts to increase below $T_{\rm H}$ and exhibits a peak at  $T_{\rm P}$.
With further decreasing $T$, the $1/T_2$ shows a minimum and then increases again below about $T_{\rm L}$. }
\end{center}
\vspace{-7mm}
\end{figure}

\begin{figure}[bt]
\begin{center}
\includegraphics[width=8.7 cm,keepaspectratio]{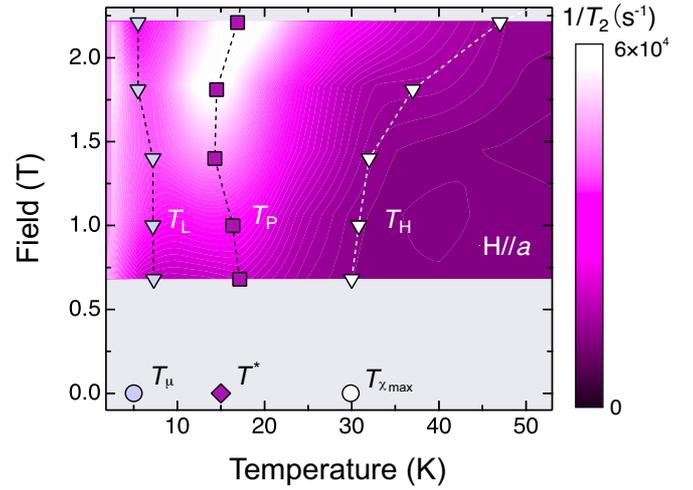} 
\caption{(Color online) The $T$-$H$ phase diagram with the contour plot of $1/T_2$ in $H\|a$.  
The phase diagram includes the $T_{\rm H}$, $T_{\rm P}$ and $T_{\rm L}$ extracted from the $T$ profile of $1/T_2$, and the other characteristic temperatures, $T_{\rm \chi,max}$, $T^{\rm *}$ and $T_{\rm \mu}$ reported previously from different experimental techniques.}
\end{center}
\vspace{-7mm}
\end{figure}

Next we discuss low-energy spin dynamics.
Figure 4 shows the $T$-dependence of $1/T_2$ measured at several different $H$.
Although the spectrum broadening was observed at low $T$, a unique $T_2$ value was obtained at each temperature on the peak position of the spectrum.
We found that there are two characteristic structures in the temperature profile of $1/T_2$.
At the lowest $H$ of 0.68 T, $1/T_2$ increases gradually below $T_{\rm H}\sim$30 K
and exhibits a broad peak at  $T_{\rm P}\sim$16 K.  The $T_{\rm H}$ agrees with the temperature below which the NMR line width shows the sudden broadening in Fig.\,3.
With further decreasing $T$, $1/T_2$ shows a minimum and then increases again below about $T_{\rm L}\sim$7 K.
We can see that the high-$T$ anomaly (the red-colored region) is substantially enhanced and broadened with increasing $H$,
 whereas  the low-$T$ anomaly (the blue-colored region) is not much affected by $H$.

In Fig.\,5, we summarize the $T$-$H$ phase diagram of UTe$_2$ in $H \| a$. 
The phase diagram includes the $T_{\rm H}$, $T_{\rm P}$ and $T_{\rm L}$ extracted from the temperature profile of $1/T_2$.
These temperatures are certainly connected to characteristic temperatures, $T_{\rm \chi,max}$, $T^{\rm *}$ and $T_{\rm \mu}$ reported previously; it has been reported that the $\chi_{b}(T)$ shows a broad maximum at $T_{\rm \chi,max}$\cite{Ran2019,Miyake2019,Knafo2019}, and resistivity, thermal-expansion and specific-heat data show anomalies at  $T^{\rm *}$ \cite{Willa,Eo,UTe2Review}.  
$T_{\rm \mu}$ is defined here as the temperature below which the $\mu$SR relaxation rate shows a rapid increase at zero-field \cite{Sunder}.
For UTe$_2$, it has been argued that the energy scale of $T_{\rm \chi,max}$ corresponds to that of the field induced
metamagnetic transition for $H||b$ \cite{Miyake2019,Knafo2019,UTe2Review}. In addition, the Kondo temperature  has been estimated to be 20$-$26 K, thus also close to $T_{\rm \chi,max}$.  
As we will discuss later, our experiments imply that long-range electronic correlations also start to develop within the U-ladder below $T_{\rm H} \simeq T_{\rm \chi,max}$.

In general, $1/T_2$ is given by the sum of electronic
and nuclear contributions, $1/T_2=(1/T_2)^{\rm el}+(1/T_2)^{\rm nu}$. 
The $(1/T_2)^{\rm nu}$ originates from the spin-spin coupling between like nuclei, and hence, should be proportional to the square root of isotope concentration. 
For UTe$_2$, $1/T_2$ values are nearly identical between the natural abundance ($7\%$) and $^{125}$Te enriched ($\sim$99\%) sample \cite{Tokunaga2019}, and hence $(1/T_2)^{\rm el}\gg(1/T_2)^{\rm nu}$.
Further, the electronic contribution consists of two terms, $(1/T_2)^{\rm el}=(1/T_2)^{\rm el}_{\|}+(1/T_2)^{\rm el}_{\bot}$ with
$(1/T_2)^{\rm el}_{\|}\propto G_{\|}(0)$ and $(1/T_2)^{\rm el}_{\bot}\propto G_{\bot}(\omega_{\rm NMR})$, where $\omega_{\rm
NMR}$ is the NMR frequency and
$G_{\alpha}(\omega)=\int_{-\infty}^{\infty}\left<h_{\alpha}(t)
h_{\alpha}(0) \right>\exp(i\omega t)dt$ is the spectral density of
the fluctuating hyperfine field, $h_{\alpha}(t)$.
Thus, $(1/T_2)^{\rm el}_{\|}$ is driven by the longitudinal
component of magnetic fluctuations at nearly zero frequency (of the order of kHz or lower), while $(1/T_2)^{\rm el}_{\bot}$ is driven by the transverse ($\bot$) components of the fluctuations at the NMR
frequency ($\sim$0.1 GHz) \cite{TokunagaURhGe}. The latter fluctuations also generate
the nuclear spin-lattice relaxation process $1/T_1$, which, however, has been found to show no significant anomaly at low-$T$ for each direction of $H$ \cite{Nakamine2019,Tokunaga2019}.
Therefore, the low-$T$ anomalies observed in $1/T_2$ are attributed to the $(1/T_2)^{\rm el}_{\|}$, thus demonstrating 
the occurrence of extremely slow and longitudinal fluctuations of the $h_{\|}(t)$ in the kilohertz or less regions along the $a$-axis.

Note that the intensity of electronic fluctuations in correlated electron systems is generally assumed to be structure-less in frequencies below megahertz region \cite{Itou,Yamamoto}. This is mostly because the energy scale of the transfer integrals and Coulomb repulsions dominating fluctuations are all in terahertz region.
However, the situation is drastically changed if the correlated electrons develop a new length scale, much longer than the atomic distance scale.
The new length scale would provide extremely slow fluctuations, in particular, near an electronic phase boundary.
Typical examples have been seen in organic materials near a metal-Mott insulator boundary \cite{Itou,Yamamoto,Urai}. 
In these materials, the electronic state becomes extremely sensitive to disorder when it approaches to the boundary. 

Theoretically, UTe$_2$ has been suggested to be located near a magnetic instability between FM and AFM order, where a long-range static order is suppressed by magnetic frustrations \cite{Xu,Ishizauka}.
Recent neutron scattering experiments also indicated 
an intra-ladder FM coupling together with an inter-ladder AFM coupling along the $b$-axis \cite{Knafo2021}.
However, for the AFM fluctuations detected by the neutron measurements, there was no signature of softening down to 2 K. 
Therefore, the slow electronic dynamics detected by $1/T_2$ would be concerned with the development of the long-range FM correlations within the U-ladder below $T_{\rm H}$.
Indeed, as seen in Fig.\,5, the anomaly below $T_{\rm H}$ is broadened and shifted to higher $T$ with increasing $H$, 
suggesting the FM character of the fluctuations along the $a$-axis. 
The observed NMR line broadening below $T_{\rm H}$ would be attributed to spatially distributed spin susceptibility induced by disturbing the long-range FM correlations with tiny amount of disorder or defects within the ladder \cite{UTe2Review,Haga}.

On the other hand, however, 
 we can not exclude the possibility that the slow fluctuations might occur in charge channel, if the system would be located near an electronic phase boundary \cite{Itou}.
It is expected that the fluctuations of the charge or the electron density at the Fermi energy cause the time-dependent modulations of the hyperfine coupling strength between U spins and Te nuclei, leading to the slow fluctuations of the $h_{\|}(t)$. 
As mentioned, the anomaly at $T^*$=$12\sim15$ K has been detected not only in $1/T_2$ but also in resistivity and thermodynamic measurements \cite{Willa,Eo,UTe2Review}. 
Certainly, there is some impact on electronic band structure around $T^*$. For the electronic state of UTe$_2$, first-principle calculations initially predict an insulating state \cite{Aoki2019}, and then the correlation effect of $f$-electrons causes an insulator-metal transition \cite{Xu,Ishizuka2019,Shick}.
Therefore, together with the magnetic instability, Fermi surface instability might also exist in UTe$_2$, where the Kondo effect would play a crucial role \cite{UTe2Review}. 
In this context, it is interesting to notice that UTe$_2$ exhibits the slow electronic dynamics resembling disordered organic materials in the vicinity of a metal-Mott insulator boundary \cite{Itou,Yamamoto,Urai}.

Finally, we note that the broad peak in $1/T_2$  might indicate the occurrence of some symmetry lowing in the electronic state below $T_{\rm P} \simeq T^*$, allowing a direct second order phase transition from the paramagnetic state to a non-unitary SC state, which is recently suggested by various experiments\cite{Jiao,Kittaka,Hayes}, but usually not allowed in the crystallographic symmetry of UTe$_2$\cite{Machida,Machida2,Kittaka,Mineev}.
Further experiments are certainly required to understand the unconventional electronic state established behind the triplet SC. 

In summary,  NMR $1/T_2$  measurements in $H\|a$ revealed slow electronic dynamics developing in the paramagnetic state of UTe$_2$.
We found that the slow fluctuations along the $a$-axis emerge below $T_{\rm \chi,max}=30-40$ K and have a peak around $T^{\rm *}=12-15$ K, followed by another rapid increase below $T_{\rm \mu}=5-7$ K. 
The behavior is concerned with the successive growth of long-range electronic correlations within the U-ladder below $T_{\rm \chi,max}$, although the origin of the two anomalies is unclear at the present stage.
At the low $T$, possibly existing tiny amounts of disorder or defects within the ladder locally disturb the long-range correlations, and develop 
an inhomogeneous electronic state associated with the low-$T$ anomaly in the $\chi_a(T)$.
We suggest that UTe$_2$ would be located on the paramagnetic side near an electronic phase boundary, where either the magnetic or Fermi-surface instability would be the origin of the characteristic fluctuations.

\subsection*{Acknowledgements} 
We are grateful for stimulating discussions with Y. Yanase, W. Knafo, J. P. Brison, J. Flouquet, M. Horvati{\'c}, M.-H. Julien, M. Hirata and V. P. Mineev. 
A part of this work was supported by JSPS KAKENHI Grant Numbers 19K03726, 19H00646, 20H00130, and 20KK0061.

\end{document}